\documentclass[aps,prd,preprint,superscriptaddress,amsfonts,amssymb,amsmath]{revtex4-1}

\usepackage{color,wrapfig,graphicx,ulem}
\usepackage[utf8]{inputenc}
\usepackage[colorlinks,linkcolor=blue]{hyperref}

\begin{document}


\title{Dark matter candidate induced by Horndeski theory: \\ dark matter halo and cosmological evolution}

\author{Jiaming Shi}
\email{2016jimshi@mails.ccnu.edu.cn}
\affiliation{Institute of Astrophysics,Central China Normal University, Wuhan 430079 ,China}
\author{Taishi Katsuragawa}
\email{taishi@mail.ccnu.edu.cn}
\affiliation{Institute of Astrophysics,Central China Normal University, Wuhan 430079 ,China}
\author{Taotao Qiu}
\email{qiutt@mail.ccnu.edu.cn}
\affiliation{Institute of Astrophysics,Central China Normal University, Wuhan 430079 ,China}

\date{\today}

\begin{abstract}
We study spherically symmetric solutions with a scalar field in the shift-symmetric subclass of the Horndeski theory. Constructing an effective energy-momentum tensor of the scalar field based on the two-fluid model, we decompose the scalar field into two components: dark matter and dark energy.
We find the dark-matter fluid is pressure-less, and its distribution of energy density obeys the inverse-square law. We show the scalar field dark matter can explain the galaxy rotation curve and discuss the time evolution of the dark matter in the cosmic background.
\end{abstract}

\maketitle



\section{Introduction}
\label{sec:intro}
The evidences for dark matter (DM) in our Universe have been accumulated so far by independent observations such as galaxy rotation curves, gravitational lensing on cluster scale, temperature fluctuations in Cosmic Microwave Background (CMB) \cite{Corbelli:1999af,Vikhlinin:2005mp,Clowe:2006eq, Ade:2015xua}.
The DM is usually considered a non-baryonic matter which mainly interacts with gravity, and DM emits little or no radiation since interactions between DM and ordinary matters are weak.
A recent study has revealed the coldness of DM through the whole cosmic history,
which implies that the DM is well-described in terms of the pressure-less fluid \cite{Kopp:2018zxp}.
Although the DM occupies about 30\% of the whole Universe, which plays an important role in the cosmological evolution,
it has not been discovered in non-gravitational experiments yet.
Therefore, we have not identified the particle nature of DM,
while various candidates are hypothesized in the models beyond the standard model of particle physics.

As an illustration, let us consider the issue of the rotational curves of the galaxies.
The rotational velocity of an object at a radial distance $r$ is given by $v \propto \sqrt{M(r)/r}$ in the Newtonian gravity,
where $M(r)$ denotes the total mass enclosed by the object's orbit.
However, the observations imply that the rotational velocity is approximately constant at a large distance $r\sim \mathcal{O}(\mathrm{kpc})$ or the larger, where there is almost no luminous matter.
It requires a DM halo with mass density $\rho (r)\propto 1/r^2$ to be introduced other than the visible matters, and the mass of the DM halo dominates the total mass of the galaxy.

Instead of the new particles beyond the standard model of particle physics,
a possible way to account for the DM is to modify the general relativity.
The modified gravity theories can introduce more degrees of freedom into the general relativity,
and the role of DM particles can be replaced with those extra degrees of freedom, to explain the galactic rotation curve \cite{Rinaldi:2016oqp,deAlmeida:2018kwq,Naik:2018mtx,Panpanich:2018cxo,Sebastiani:2016ras}.
Generally speaking, the additional degrees of freedom, which are often rewritten in terms of new dynamical fields, change the gravitational interaction.
Such new fields directly affect and contribute to the spacetime as new matters, literally as the DM.
In other words, the modified gravity can allow us to discuss the DM as a gravitational phenomenon which is unexplainable in the framework of the general relativity and ordinary matters.
In this work, we cast a scalar field induced by the modified gravity as the DM fluid and investigate its nature.

We consider a simple case that the DM is rephrased with the scalar field in the scalar-tensor theory.
For instance, k-essence theory has been studied in \cite{ArmendarizPicon:2005nz} to explain the rotation curve.
On the other hand, static and inhomogeneous configurations of the scalar field usually lead to the anisotropic pressure in the scalar-tensor theory.
Actually, several works have suggested that a fluid with pressure can generate a halo and result in a flat rotation curve \cite{Nucamendi:2000jw,Bharadwaj:2003iw}.
However, these results are contradictory with the cosmological observation \cite{Kopp:2018zxp},
which implies that the DM fluid is pressure-less.

Now we take a look at a wider framework of the scalar-tensor theory,
the generalized Galileon theory \cite{Deffayet:2009wt,Deffayet:2009mn,Kobayashi:2011nu},
which was found to be identical to the Horndeski theory \cite{Horndeski:1974wa} established in 1974.
As a new scalar-tensor theory, the Horndeski theory introduces four arbitrary functions of scalar $\phi$, and the scalar field generates the fifth force which modifies the dynamics of the general relativity.
Although many works on Horndeski theory have revealed the cosmological solutions (see \cite{Kobayashi:2019hrl} for a review),
it is necessary to check the applicability of Horndeski theory to the sub-cosmological scale physics,
which help us comprehend the Horndeski theory as the theory of gravitation in the Universe.
In 2012, the first nontrivial static black hole solution was found in a subclass of Horndeski theory \cite{Rinaldi:2012vy},
and later, a time-dependent black hole solution was done by \cite{Babichev:2013cya}.
Further studies on the Horndeski theory at local scale, including new spacetime solutions \cite{Minamitsuji:2013ura,Anabalon:2013oea,Kobayashi:2014eva,Maselli:2015yva,Babichev:2015rva,Babichev:2017guv,Sebastiani:2018hak},
self-tuning issues \cite{Appleby:2015ysa,Babichev:2016kdt,Babichev:2017lmw},
scalar hair \cite{Ogawa:2015pea,Hui:2012qt,Sotiriou:2013qea,Tattersall:2018map,Babichev:2016rlq},
and so on have been under discussion.

On the other hand, the recent gravitational wave event GW170817 and its seemingly electromagnetic counterpart GRB170817A have given stringent constraints on the modified gravity theories \cite{Baker:2017hug,Creminelli:2017sry,Sakstein:2017xjx,Ezquiaga:2017ekz}.
Therefore, describing the DM in the framework of the modified gravity theories is more challenging under such constraints\cite{Diez-Tejedor:2018fue}.
For example, the earlier work \cite{Casalino:2018mna} suggests that the consistency between the DM behavior in cosmological evolution and the constraints on the speed of the gravitational wave is violated in some subclasses of the Horndeski theory. It is also interesting to find a more optimized model which can explain the DM halo and cosmic evolution as well as survive after GW170817.

In this paper, we will improve the model in \cite{Rinaldi:2012vy} to study new black hole solutions
and discuss an application to describe the DM halo.
We employ a spherically symmetric solution to study the rotational curve.
In order to address the anisotropic pressure of the scalar field, we make use of the two-fluid model \cite{Letelier:1980mxb,Bayin:1985cd,Dey:2013vaa} and study the cosmic evolution.
Dividing the energy-momentum tensor of the scalar field into two perfect fluids,
we show that one fluid corresponding to DM can be pressure-less and another works as Dark Energy (DE).
This paper is organized as follows.
In Sec.~\ref{sec:horndeski}, we provide an action possessing the shift symmetry of the scalar field $\phi$ and utilize this action to study a spherically symmetric black hole solution.
Moreover, we consider the application of this solution to explain the galactic rotation curve.
In Sec.~\ref{TFModel},
we apply the two-fluid model to study the anisotropic fluid composed by the scalar field and obtain a pressure-less DM fluid.
Next in Sec.~\ref{DMBcos}, we discuss the behavior of the DM fluid in cosmological evolution,
to find that it is consistent with the result in galaxy scale.
We will also face our model to the constraint on the sound speed squared of the gravitational waves given by GW170817 and GRB170817A.
Finally, we conclude our result in Sec.~\ref{sec:conclusion}.


\section{Shift-Symmetric Horndeski Theory and its Black Hole Solution}
\label{sec:horndeski}

In this section, we briefly review the Horndeski theory and introduce a particular subclass by imposing symmetries.
Assuming specific forms of the arbitrary functions in the Horndeski theory, we demonstrate the existence of a spherically symmetric black hole solution.
Moreover, we define the energy-momentum tensor of the scalar field and analyze the energy density and pressure of the scalar field based on the fluid description.


\subsection{The Horndeski theory}
The Horndeski theory is described by the following action:
\begin{align}
\label{horndeski_action}
S= \sum^{5}_{i=2} \int d^{4}x \sqrt{-g} \mathcal{L}_{i}\, ,
\end{align}
where $\mathcal{L}_{i}$ are defined as follows:
\begin{align}
\mathcal{L}_{2}=&G_{2} (\phi, X) \, ,
\\
\mathcal{L}_{3}=& - G_{3} (\phi, X) \Box \phi \, ,
\\
\mathcal{L}_{4}=&G_{4} (\phi, X) R + G_{4X} \left[ (\Box \phi)^{2} - ( \nabla_{\mu} \nabla_{\nu} \phi)^{2} \right]\, ,
\\
\mathcal{L}_{5}=&G_{5} (\phi, X) G_{\mu \nu} \nabla^{\mu} \nabla^{\nu} \phi - \frac{G_{5X}}{6}
\left[ (\Box \phi)^{3} - 3 (\Box \phi) (\nabla_{\mu} \nabla_{\nu} \phi)^{2} + 2 (\nabla_{\mu} \nabla_{\nu} \phi)^{3} \right] \, .
\end{align}
$G_{i}(\phi, X)$ are arbitrary functions of the scalar field $\phi$ and its kinetic term $X=-(\partial_{\mu} \phi)^{2}/2$, $R$ is the Ricci scalar, and $G_{\mu \nu}$ is the Einstein tensor. We choose the convention such that $M_p^{-2} \equiv 8\pi G=1$, and use the notation such as $f_X\equiv\partial f/\partial X$, $f_{\phi}\equiv\partial f/\partial \phi$ to describe the derivatives of a function $f(\phi, X)$ with respect to $\phi$ and $X$. Note that by choosing specific forms of the above functions $G_{i}(\phi, X)$, the Horndeski theory turns to coincide with the various modified gravity theories, including the general relativity.

Varying the action \eqref{horndeski_action}, we obtain
\begin{align}
\label{variation}
\delta S
= \sqrt{-g} \left[ \sum^{5}_{i=2} \mathcal{G}^{i}_{\mu \nu} \delta g^{\mu \nu}
+ \sum^{5}_{i=2} \left(\mathcal{P}^{i}_{\phi} - \nabla^{\mu}\mathcal{J}^{i}_{\mu} \right) \delta \phi \right] \, ,
\end{align}
up to the total derivatives. The concrete expressions of $\mathcal{G}^{i}_{\mu \nu}$, $\mathcal{P}^{i}_{\phi}$, and $\mathcal{J}^{i}_{\mu}$ are listed in the appendix of Ref.~\cite{Kobayashi:2014eva}. Then, the equation of motion for both metric and the scalar field in the Horndeski theory are symbolically expressed as
\begin{align}
\label{horndeski_eom1}
\sum^{5}_{i=2} \mathcal{G}^{i}_{\mu \nu} = 0
\, ,
\end{align}
and
\begin{align}
\label{horndeski_eom2}
\sum^{5}_{i=2}\mathcal{P}^{i}_{\phi} - \sum^{5}_{i=2} \nabla^{\mu}\mathcal{J}^{i}_{\mu}  = 0 \, ,
\end{align}
respectively. It is known that the equations of motion Eqs.~\eqref{horndeski_eom1} and \eqref{horndeski_eom2} are of the second order of derivatives, which allows us to avoid the Ostrogradsky instability.


\subsection{Shift-symmetric model}

We consider the following subclass of the Horndeski theory:
\begin{align}
\label{action}
\mathcal{L}_2=G_{2}(X) \, ,\, \mathcal{L}_4=G_{4} (X) R + G_{4X} \left[ (\Box \phi)^{2} - ( \nabla_{\mu} \nabla_{\nu} \phi)^{2} \right] \, , \, \mathcal{L}_3=\mathcal{L}_5=0\, .
\end{align}
The above model possesses the shift symmetry, $\phi \rightarrow \phi + c$ where $c$ is a constant, and the $Z_{2}$ symmetry, $\phi \rightarrow - \phi$. These two symmetries allow only $G_{2}$ and $G_{4}$ in the Lagrangian, and these functions depend only on $X$. The above subclass was introduced and examined by earlier works \cite{Minamitsuji:2013ura,Anabalon:2013oea,Kobayashi:2014eva,Maselli:2015yva,Babichev:2015rva}.

In this case, Eq. (\ref{horndeski_eom1}) is written as
\begin{align}
\label{EoM}
\mathcal{E}_{\mu\nu}\equiv\mathcal{G}_{\mu\nu}^{(2)} + \mathcal{G}_{\mu\nu}^{(4)} = 0 \, .
\end{align}
Here, each part is given as (also see, \cite{Kobayashi:2011nu})
\begin{eqnarray}
\label{E2}
\mathcal{G}_{\mu\nu}^{(2)}
&=&-\frac{1}{2} G_{2X} \nabla_{\mu} \phi \nabla_{\nu} \phi - \frac{1}{2} G_{2} g_{\mu\nu}, \\
\label{E4}
\mathcal{G}_{\mu\nu}^{(4)}&=&G_{4}G_{\mu\nu}
- \frac{1}{2}G_{4X}R \nabla_{\mu} \phi \nabla_{\nu} \phi
- \frac{1}{2} G_{4XX} \left[
(\Box\phi)^{2}
- (\nabla_{\alpha} \nabla_{\beta} \phi)^{2}
\right]
\nabla_{\mu} \phi \nabla_{\nu} \phi
\nonumber \\
&&
- G_{4X} \Box \phi \nabla_{\mu} \nabla_{\nu} \phi
+ G_{4X} \nabla_{\lambda} \nabla_{\mu} \phi \nabla^{\lambda} \nabla_{\nu} \phi
+ 2\nabla_{\lambda}G_{4X}\nabla^{\lambda}\nabla_{(\mu} \phi \nabla_{\nu)} \phi
- \nabla_{\lambda}G_{4X} \nabla^{\lambda} \phi \nabla_{\mu} \nabla_{\nu} \phi
\nonumber \\
&&
+ g_{\mu\nu} \left\{
G_{4XX} \nabla_{\alpha} \nabla_{\lambda} \phi \nabla_{\beta} \nabla^{\lambda} \phi \nabla^{\alpha} \phi \nabla^{\beta} \phi
+ \frac{1}{2}G_{4X}\left[(\Box\phi)^{2}-(\nabla_{\alpha}\nabla_{\beta}\phi)^{2}\right]\right\}
\nonumber \\
&&
+ 2 \left[
G_{4X}R_{\lambda(\mu}\nabla_{\nu)} \phi \nabla^{\lambda} \phi
- \nabla_{(\mu} G_{4X} \nabla_{\nu)} \phi \Box \phi
\right]
-g_{\mu\nu}\left[
G_{4X}R^{\alpha\beta}\nabla_{\alpha}\phi\nabla_{\beta}\phi
-\nabla_{\lambda} G_{4X} \nabla^{\lambda} \phi \Box \phi
\right]
\nonumber \\
&&
+G_{4X} R_{\mu \alpha \nu \beta} \nabla^{\alpha} \phi \nabla^{\beta}\phi
- G_{4XX}\nabla^{\alpha}\phi\nabla_{\alpha}\nabla_{\mu}\phi\nabla^{\beta}\phi\nabla_{\beta}\nabla_{\nu}\phi
\, ,
\end{eqnarray}
where the parenthesis $()$ expresses the symmetric part with two indices swapped, $A_{(\mu \nu)} = \frac{1}{2} (A_{\mu \nu} + A_{\nu \mu})$.
Corresponding to the shift symmetry $\phi\rightarrow\phi+c$,
we have a Noether current $\mathcal{J}^{\mu}$,
\begin{eqnarray}
\label{current}
\mathcal{J}^{\mu}&\equiv&\frac{\delta(\mathcal{L}_2+\mathcal{L}_4)}{\delta \nabla_{\mu}\phi} \, \nonumber\\
&=&
-\mathcal{\mathit{G}}_{2X} \nabla^{\mu} \phi
+ 2G_{4X} G^{\mu\nu} \nabla_{\nu} \phi
- G_{4XX} \left[(\Box \phi)^{2} - (\nabla_{\mu} \nabla_{\nu} \phi)^{2} \right] \nabla^{\mu} \phi
\nonumber \\
&&
-2G_{4XX} \nabla^{\mu} X \Box \phi
+2G_{4XX} \nabla^{\nu} X \nabla^{\mu} \nabla_{\nu} \phi \, .
\end{eqnarray}
Note that the equation of motion with respect to the scalar field is rephrased as the current conservation law:
\begin{align}
\label{current_conservation}
\nabla_{\mu}\mathcal{J}^{\mu}=0\, .
\end{align}

 In the following analysis, we employ a toy model with the specific choice of $G_{2}(X)$ and $G_{4}(X)$:
\begin{align}
\label{model}
\begin{split}
G_{2}(X) &= -2\Lambda + 2\eta X^p \,, \\
G_{4} (X) &= \zeta+\beta X^p \, ,
\end{split}
\end{align}
where $\Lambda$, $\eta$, $\xi$, $\beta$, and $p$ are parameters. Note that for $p=1$, the above model reproduces the so-called non-minimal derivative coupling $G^{\mu\nu}\nabla_{\mu}\phi\nabla_{\nu}\phi$, as in \cite{Maselli:2015yva}
\begin{align}\label{nonDC}
S
=&
\int d^{4} x \sqrt{-g} \left[
-2\Lambda
+ 2\eta X
+ (\zeta+\beta X)R
+ \beta \left[
(\Box\phi)^{2}
- (\nabla_{\mu}\nabla_{\nu} \phi)^{2} \right]
\right]
\nonumber \\
=&
\int d^{4} x \sqrt{-g} \left[\zeta R -2\Lambda + 2\eta X + \beta G^{\mu\nu}\nabla_{\mu}\phi\nabla_{\nu}\phi \right] \, ,
\end{align}
where we have used the following equation defining Riemann curvature tensor,
\begin{align}
\label{eq1}
&\nabla_{\mu}\nabla_{\nu}\nabla^{\mu}\phi-\nabla_{\nu}\nabla_{\mu}\nabla^{\mu} \phi
= R^{\rho}_{\ \nu} \nabla_{\rho} \phi\,.
\end{align}
The spherically symmetric black hole solutions have been studied by \cite{Rinaldi:2012vy,Babichev:2013cya,Minamitsuji:2013ura,Anabalon:2013oea} in the case where $p=1$, and thus, our model can be considered a generalization of the model which the earlier works have employed.


\subsection{Spherically symmetric system}

In order to analyze the galaxy rotation curve, we first consider the static and spherically symmetric spacetime solutions in the shift-symmetric model Eq.~\eqref{model}, which gives us approximated descriptions of the spacetime around the galaxy. We start with the general form of the static and spherically symmetric spacetime:
\begin{align}
\label{m2}
ds^{2}=-h(r)dt^{2}+\frac{1}{f(r)}dr^{2}+r^{2}d\Omega^{2}.
\end{align}
For the scalar field $\phi$, we also assume the scalar field is static and the spherically symmetric, that is, $\phi=\phi(r)$.

The nonvanising components of $\mathcal{E}_{\mu\nu}$ and $\mathcal{J}^\mu$ are given by
\begin{eqnarray}
\label{Ett}
\mathcal{E}_{tt}&=&\frac{8hfXX'G_{4XX}}{r}+\frac{4fhXG_{4X}}{r^{2}}+\frac{4f'hXG_{4X}}{r}+\frac{4fhX'G_{4X}}{r} \nonumber \\
&&-\frac{2fhG_{4}}{r^{2}}+h G_{2}-\frac{2f'hG_{4}}{r}+\frac{2G_{4}h}{r^{2}}\,, \\
\label{Err}
\mathcal{E}_{rr}&=&-\frac{8G_{4XX}X^{2}}{r^{2}}-\frac{8h'G_{4XX}X^{2}}{hr}-\frac{8G_{4X}X}{r^{2}}-\frac{8h'G_{4X}X}{hr}+\frac{2XG_{2X}}{f}+\frac{4G_{4X}X}{fr^{2}} \nonumber \\
&&-\frac{2G_{4}}{r^{2}}+\frac{2G_{4}h'}{hr}-\frac{G_{2}}{f}-\frac{2G_{4}}{fr^{2}}\,, \\
\mathcal{E}_{\theta\theta}&=&-\frac{2XX'G_{4XX}fh'r^{2}}{h}-\frac{2XG_{4X}fh''r^{2}}{h}+\frac{XG_{4X}fh'^{2}r^{2}}{h^{2}}
\nonumber \\
&&-4XX'G_{4XX}fr-\frac{2XG_{4X}fh'r}{h}-\frac{XG_{4X}f'h'r^{2}}{h}-2XG_{4X}f'r
\nonumber \\
&&-\frac{X'G_{4X}fh'r^{2}}{h}+\frac{G_{4}fh''r^{2}}{h}-\frac{G_{4X}fh'^{2}r^{2}}{2h^{2}}\nonumber \nonumber \\
&&-2X'G_{4X}fr+\frac{G_{4}fh'r}{h}+\frac{G_{4}f'h'r^{2}}{2h}+G_{4}f'r-G_{2}r^{2}\,, \\
\mathcal{E}_{\varphi\varphi}&=&\sin^2\theta\mathcal{E}_{\theta\theta}\,, \\
\label{Jr}
\mathcal{J}^{r}
&=& \frac{f\phi'}{r^{2}h}\left\{ -(r^{2}G_{2X}+2G_{4X})h+2(G_{4X}+2XG_{4XX})(rh)'f\right\}\,,
\end{eqnarray}
where the prime expresses the derivative with respect to $r$. Note that we have the Noether current $\mathcal{J}^\mu=(0,\mathcal{J}^r,0,0)$ under the assumption of static field configuration. One can find that the conservation law leads to $\partial_r (\sqrt{h/f}\mathcal{J}^rr^2)=0$, and a solution is given by $\sqrt{h/f}\mathcal{J}^rr^2=C$ where $C$ is an integration constant. If we want to get a black hole solution, there should exist a horizon $r=r_h$, where $f(r_h)=0$. Therefore from Eq. (\ref{Jr}), one can find that this will lead to a vanishing current, namely $\mathcal{J}^r=0$. This can be viewed as a specific solution of the continuity equation (\ref{current_conservation}), although not necessary at the horizon. \footnote{For non-rotating black hole solution, $h(r_h)=f(r_h)=0$ while $h(r_h)/f(r_h)$ is still finite, so our solution does not change. A well-known example is the Schwartzschild solution where $h(r)=f(r)=1-2GM/r$.}

Imposing the vanishing current in terms of Eq.~\eqref{current_conservation} and using Eq. \eqref{EoM}, we have four equations
\begin{align}
\label{eomXp}
\mathcal{J}^{\mu}=0 \, , \
\mathcal{E}_{tt}=0 \, , \
\mathcal{E}_{rr}=0 \, , \
\mathcal{E}_{\theta\theta}=0
\end{align}
where the first equation is a specific solution of the continuity equation (\ref{current_conservation}). For the case where $\eta\beta>0$, Eqs.~\eqref{eomXp} give the following solutions of $h(r)$, $f(r)$, and $\phi(r)$:
\begin{align}
\label{sol2}
h(r) =&
\frac{\beta(2p-1)^{2}(\Lambda\beta+\zeta\eta)^{2}}{4\zeta^{2}p^{2}\sqrt{\eta\beta}\eta^{2}r} \arctan(\frac{r\eta}{\sqrt{\eta\beta}})
-\frac{(2p-1)(\eta r^{2}+6\beta p-3\beta)}{6\zeta\eta p^{2}}\Lambda
\nonumber \\
&
- \frac{\beta(2p-1)^{2}(-\eta r^{2}+3\beta)}{12\zeta^{2}\eta^{2}p^{2}}\Lambda^{2}
- \frac{\mu}{r}
+ \frac{\eta r^{2}}{12\beta p^{2}}
+ \frac{1}{p}
-\frac{1}{4p^{2}}
\,, \\
\label{sol3}
f(r)=& \frac{h(r)}{w(r)}
\,, \\
\label{sol4}
w(r)=&\frac{(2p-1)\left(2\Lambda\beta pr^{2}-\Lambda\beta r^{2}-\zeta\eta r^{2}-2\zeta\beta p\right)^{2}}{4\zeta^{2}p^{2}\left(\eta r^{2}+\beta\right)^{2}}
\,, \\
\label{sol}
\phi'(r)^{2}=&
-2\left(\frac{2\Lambda\beta pr^{2}-\Lambda\beta r^{2}+\zeta\eta r^{2}-2\zeta\beta p+2\zeta\beta}{2(2p-1)(\eta r^{2}+\beta)\beta}\right)^{\frac{1}{p}}f(r)^{-1}
\, ,
\end{align}
where a parameter $\mu$ in Eq.~\eqref{sol2} is an integration constant. If $\eta\beta<0$, one can find a similar solution by replacing
$\sqrt{\eta \beta}\rightarrow \sqrt{-\eta \beta}$ and $\arctan(r\eta/\sqrt{\eta\beta}) \rightarrow \mathrm{arctanh}(r\eta/\sqrt{-\eta\beta})$ in Eq.\eqref{sol2} while keeping Eqs.~\eqref{sol3}, \eqref{sol4}, and \eqref{sol} unchanged. As an example, our generalized solutions can surely reproduce the black hole solution derived in the earlier work \cite{Rinaldi:2012vy} when we choose a specific set of parameters, $p=1$, $\Lambda=0$, $\eta=1/2$, $\zeta=1/2$, and $\beta=z_0/2$:
\begin{align}
\label{nc}
h(r)&=\frac{3}{4}+\frac{r^{2}}{12z_0}-\frac{\mu}{r}+\frac{\sqrt{z_0}}{4r}\arctan\left(\frac{r}{\sqrt{z_0}}\right)\,,\\
f(r)&=\frac{4\left(r^{2}+z_0\right)^{2}h(r)}{\left(r^{2}+2z_0\right)^{2}}\,,\\
\phi'(r)^{2}&=-\frac{r^{2}\left(r^{2}+2z_0\right)^{2}}{4z_0\left(r^{2}+z_0\right)^{3}h(r)}\,.
\end{align}

As a side remark, we note that this solution cannot be applied to the specific parameter choice $p=1/2$, since $\phi'(r)^2$ diverges in Eq. \eqref{sol}. Using the original continuity equation $\nabla_\mu J^\mu= 0$ one can get another class of solution:
\begin{eqnarray}
\label{F12}
h(r)&=&\frac{\Sigma}{(\eta r^{2}+\beta)^{2}}\,,\\
\label{G12}
f(r)&=&\frac{\left(\eta r^{2}+\beta\right)\left(\Lambda r^{2}-\zeta\right)}{\zeta\left(3 \eta r^{2}-\beta\right)}\,,\\
\label{phi12}
\phi'(r)^{2}&=&\frac{8\zeta r^{4}(-9\Lambda\eta^{2}r^{4}+4\Lambda\beta\eta r^{2}+6\zeta\eta^{2}r^{2}+\Lambda\beta^{2}-6\zeta\beta\eta)^{2}}{(\eta r^{2}+\beta)^{3}(-3\eta r^{2}+\beta)^{3}(\Lambda r^{2}-\zeta)}\,.
 \end{eqnarray}
where $\Sigma$ represents an integration constant. It is obvious that the solution gives the horizon at $r_h=\sqrt{\zeta/\Lambda}$, but there is no infinite redshift surface which causes $h(r)=0$, so it shows significant difference from a black hole solution.

Next, we look further into the solutions Eqs.~\eqref{sol2}, \eqref{sol3}, \eqref{sol4}, and \eqref{sol}. Using the definition $X=-f(r)\phi'(r)^2/2$, one finds
\begin{align}
\label{Xsol}
X^p=\frac{2\Lambda\beta pr^{2}-\Lambda\beta r^{2}+\zeta\eta r^{2}-2\zeta\beta p+2\zeta\beta}{2(2p-1)(\eta r^{2}+\beta)\beta}\, .
\end{align}
For simplicity, we focus on the specific case where $\beta \Lambda + \zeta \eta = 0$, therefore the first term of Eq.~\eqref{sol2} could be canceled. Defining $z \equiv \beta / \eta = - \zeta / \Lambda$, the black hole solution is expressed by the following simple form:
\begin{eqnarray}
\label{FG}
h(r)&=&1-\frac{\mu}{r}-\frac{r^2}{3(-z)} \, , \\
\label{FG2}
f(r)&=&(2p-1)^{-1}h(r) \, , \\
\label{FG3}
X^p&=&\frac{(p-1)\Lambda}{(2p-1)\eta} \,.
\end{eqnarray}
It is remarkable that our model produces a constant solution for $X$. Hence, as in the equation of motion \eqref{E4}, the coefficient $G_4$ in front of the Einstein tensor $G_{\mu\nu}$ is constant, which would be absorbed into the redefinition of the Planck scale. This result implies that our solution does not break Einstein's equivalence principle.

Hereafter, we assume $\zeta, \eta, \Lambda>0$, $\beta<0$, and $p>1$.
In those parameter regions,
$h(r)$ takes the similar form to the Schwarzschild-de Sitter solution with positive cosmological constant $1/(-z) = \Lambda/\zeta>0$ as $X>0$ and $X^p>0$.
However, it does {\it not} mean that our model is as simple as GR plus a cosmological constant.
This is because the nontrivial coupling between the scalar field and gravity still appears in the Einstein Equations, and the energy-momentum tensor, as we will show soon, has $r$-dependence and can play the role of DM.
A similar case can be found in the vacuum solution of ghost condensate field in cosmology \cite{ArkaniHamed:2003uy,ArkaniHamed:2003uz}.

\subsection{Energy-Momentum Tensor of Scalar Field}
The equation of motion \eqref{EoM} can be rewritten in a similar form of the Einstein equation, given as
\begin{align}
\label{ge}
G_{\mu\nu}=T^{eff}_{\mu\nu}\,,
\end{align}
 Here, $T^{eff}_{\mu\nu}$ represents the energy-momentum tensor of the scalar field. Since for spherical symmetric solutions one has $G_{r}^r\neq G_{\theta}^{\theta}\neq G_{\varphi}^\varphi$, namely $G_{i}^j$ is anisotropic in these three directions, one will expect the energy-momentum $T^{eff}_{\mu\nu}$ to be the same. Therefore we write the effective energy momentum tensor $T^{eff}_{\mu\nu}$ as \cite{Wald:1984rg}:
\begin{align}
\label{Tmunu0}
T^{eff}_{\mu\nu}= \rho e^0_\mu e^0_\nu+p_{\|} e^1_\mu e^1_\nu+p_{\perp} e^2_\mu e^2_\nu+p_{\perp} e^3_\mu e^3_\nu\,.
\end{align}
Here, the tetrads $e^a_\mu$ satisfy $g^{\mu\nu}e^a_\mu e^b_\nu=\eta^{ab}$, where $\eta^{ab}=\mathrm{diag}(-1,1,1,1)$ is the Minkowski spacetime metric, and $e^a_\mu=\mathrm{diag}(\sqrt{-g_{00}}, \sqrt{g_{11}}, \sqrt{g_{22}}, \sqrt{g_{33}})$. The energy density and the pressures of scalar field are defined as follows,
\begin{align}
\label{rhop}
\rho&\equiv -T_0^0=\frac{2(p-1)}{(2p-1)r^2}+\frac{1}{(2p-1)(-z)} \,, \\
\label{rhop2}
p_{\|}&\equiv T_1^1=-\frac{2(p-1)}{(2p-1)r^2}-\frac{1}{(2p-1)(-z)}\,,  \\
\label{rhop3}
p_{\perp}&\equiv T_2^2=T_3^3=-\frac{1}{(2p-1)(-z)}\,.
\end{align}

One can find that the second term in Eq.~\eqref{rhop} and \eqref{rhop2}, as well as the term in Eq. \eqref{rhop3}, are behaving like a cosmological constant with isotropic pressure.
So, this part could be naively viewed as dark energy (DE).
Moreover, the first term in Eq.~\eqref{rhop} is proportional to $r^{-2}$.
According to the analysis in \cite{ArmendarizPicon:2005nz}, this behavior of DM can explain the gravitational rotation curve.
However, it is well-known that the cold DM needs vanishing pressure in all directions.
Although the corresponding terms in $p_{\perp}$ can be viewed as vanishing, that in $p_{\|}$ is obviously non-vanishing, making this term fail to be explained as DM.
To make out of this dilemma, in the next section, we will take a different view of this model.


\section{Anisotropic fluid and two-fluid model}
\label{TFModel}
\subsection{DM and DE fluids}
Due to the anisotropy of the energy momentum tensor, $p_{\|}-p_{\perp}\neq0$, in Eq.(\ref{Tmunu0}), it can be decomposed into the two non-interacting perfect fluids sourcing the spacetime structure of solution Eqs.~\eqref{FG}\eqref{FG2}\eqref{FG3},  expressed as
\begin{align}\label{twofluids}
 T_{\mu\nu}=(p_1+\rho_1)u_\mu u_\nu+p_1g_{\mu\nu}+(p_2+\rho_2)v_\mu v_\nu+p_2g_{\mu\nu}\,.
\end{align}
Here $\rho_i$ and $p_i$ ($i=1, 2$) denote the energy density and pressure of each perfect fluid, $v^\mu, u^\mu$ are timelike 4-velocities of each component of two-fluid system, i.e., $u_\mu u^\mu = v_\mu v^\mu=-1$, and the anisotropy vanishes if $u^\mu=v^\mu$.

Using Eqs.(\ref{Tmunu0}) and (\ref{twofluids}), the energy density $\rho$ and the pressures $p_{\|}$, $p_{\perp}$ are given by \cite{Letelier:1980mxb,Bayin:1985cd,Dey:2013vaa}
\begin{align}
\label{relations}
\begin{split}
\rho&=\frac{1}{2}(\rho_1-p_1+\rho_2-p_2)+\frac{1}{2}\sqrt{(\rho_1+p_1+\rho_2+p_2)^2+4(\rho_1+p_1)(\rho_2+p_2)(K^2-1)}\,,  \\
p_{\|}&=-\frac{1}{2}(\rho_1-p_1+\rho_2-p_2)+\frac{1}{2}\sqrt{(\rho_1+p_1-\rho_2-p_2)^2+4(\rho_1+p_1)(\rho_2+p_2)K^2} \,, \\
p_{\perp}&=p_1+p_2 \,,
\end{split}
\end{align}
where $K\equiv v_\mu u^\mu<0$ since $v^\mu, u^\mu$ are timelike vectors. From equations (\ref{relations}), we obtained the analytical solution:
\begin{align}
\label{solutions}
\begin{split}
\rho_1&=\frac{(xn+1)r^2-4(p-1)z}{(x-1)(2p-1)r^2z}\,, \\
\rho_2&=\frac{-x[(n+1)r^2-4(p-1)z]}{(x-1)(2p-1)r^2z}\,,  \\
p_1&=\frac{\rho_1}{n} \,, \\
p_2&=\frac{\rho_2}{xn} \,,  \\
K&=\pm \frac{r^2(n+1)(xn+1)-2(p-1)z(xn+n+2)}{\sqrt{(n+1)(xn+1)[(n+1)r^2-4(p-1)z][r^2(xn+1)-4(p-1)z]}} \,.
\end{split}
\end{align}
Here, since there are fewer variables than the equations in (\ref{relations}), and thus the solution cannot be uniquely determined,
we introduce two free parameters $x$ and $n$.
One can find that the equation-of-state parameters associated with the two fluids are characterized by the above two parameters,
such as $w_{1} \equiv 1/n$ and $w_{2} \equiv 1/(xn)$.

Let us assume that the second fluid $(\rho_{2}, p_{2})$ behaves as a dust, and we take the limit $x \rightarrow \pm \infty$ where $w_{2} \rightarrow 0$.
Under this condition, the density profiles of the two fluids are given by
\begin{align}
\rho_{1} \rightarrow \frac{n}{(2p-1)z} \, , \
\rho_{2} \rightarrow \frac{4(p-1)}{(2p-1) r^2} - \frac{n+1}{(2p-1)z} \, .
\label{density_profiles}
\end{align}
One finds that the energy density for the first fluid $(\rho_{1}, p_{1})$ is always constant,
while that for the second fluid shows the inverse-square law with respect to the radial coordinate $r$.
Because $z$ is negative, we find a constraint on $n$ to obtain the positive energy densities:
\begin{align}
-1 \leq n \leq 0
\, .
\end{align}
Therefore, the first fluid can behave as the cosmological constant ($w_{1} = -1$) or the phantom ($w_{1} < -1$) in our setup.
Hereafter, we consider the simplest case to induce the $\Lambda$CDM-like model by setting $n\rightarrow-1$ and $x\rightarrow\pm\infty$,
where the density profiles of the two fluids are determined as follows:
\begin{align}
\label{solutions20}
\begin{split}
\rho_1\rightarrow\frac{1}{(2p-1)(-z)}\,, \,
\rho_2\rightarrow\frac{4(p-1)}{(2p-1)r^2}\,, \,
p_1\rightarrow-\rho_1 \,,\,
p_2\rightarrow0 \,,  \,
K\rightarrow\pm \infty \,.
\end{split}
\end{align}

In Eqs.~\eqref{solutions20}, $\rho_1$ can be read off as the DE density, while $\rho_2$ can be \textcolor{blue}{as} the DM density
since the equation-of-state parameter $w_{DM}\equiv p_2/\rho_2=0$.
That is, we can regard the first fluid $(\rho_{1}, p_{1})$ as the DE fluid as the second one $(\rho_{2}, p_{2})$ as the DM fluid in this scenario.
Note that the total mass enclosed by the orbit is symbolically written as $m(r)=\int [\rho_2(r)+\rho_e(r)] dV$,
where $\rho_e$ is the average density of environment of the ordinary matter rather than the scalar field,
and $\rho_1$ should be so small compared to $\rho_e$ that it can be negligible.

We further explain the physical meaning of $K$ reaching infinity. In general, the 4-velocity can be decomposed like $u^\mu=(Z^\mu+\bar{u}^\mu)/\sqrt{1-\bar{u}^\mu\bar{u}_\mu}$, where $Z^u$ is the 4-velocity of observer at rest in the coordinate system, and $\bar{u}^\mu$ is spacelike satisfying $Z_\mu\bar{u}^\mu=0$.  Here  the 4-velocity of observer can be $v^\mu$, thus this observer of  4-velocity  $v^\mu$ measures a particle of 4-velocity $u^\mu=\gamma_\lambda v^\mu+\gamma_{\lambda} \lambda^i\delta_i^\mu$, $i=1,2,3$, where $\gamma_\lambda=1/\sqrt{1-\lambda^i\lambda_i} \geq1$ is the Lorentz factor. Thus  $K=v_\mu u^\mu$ can be rewritten by the 3-velocity $\lambda^i$ which is measured by the observer of  4-velocity $v^\mu$, that is
\begin{align}
\label{Kurelation}
K=v_\mu u^\mu=-\frac{1}{\sqrt{1-\lambda^i\lambda_i}}\,.
\end{align}
The right-hand side of the Eq.~\eqref{Kurelation} can be regarded as the Lorentz factor $\gamma_\lambda=1/\sqrt{1-\lambda^i\lambda_i} \geq1$. Therefore, if $n\rightarrow-1$ and $K\rightarrow-\infty$, the relative velocity between the DM and DE fluids is the speed of light.

\subsection{Facing observational constraints on DM fluids}
{\it Galaxy rotation curve.}
Without the loss of generality of our model, we can choose $\zeta=1$ for simplicity. To study the rotation curve, we need to know the velocity of object travelling around the galaxy, which is given by the geodesic equation in the Newtonian limit,
\begin{align}\label{speed2}
 v^2\equiv r \Gamma_{00}^r\simeq \frac{G_N m(r)}{r}-\frac{\Lambda}{3(2p-1)}r^2\,,
\end{align}
where $G_N=G/(2p-1)$.
$\Gamma_{00}^r$ is the ($t,t,r$) component of Christoffel symbol $\Gamma_{\mu\nu}^\lambda$,
and the parameter $m(r)$ is related to the mass of DM and ordinary matter enclosed by the orbit.

Based on the previous analysis in the two-fluid model,
when we consider $\Lambda$ in Eq.~\eqref{model} as the cosmological constant,
it is so small that we can neglect its effect at the galactic scales.
The Eq. \eqref{speed2} can approximate
\begin{align}\label{dmdisk2}
v=\sqrt{\frac{G_{N}m(r)}{r}} \, , \ m(r)=\int \rho dV=\frac{4\pi}{3}(3\rho_2+\rho_e) r^{3} \, ,
\end{align}
and we have two cases of the velocity of test particle:
\begin{align}
\label{twocases}
\left\{
\begin{array}{ll}
v\propto r \ &\mbox{when} \quad \rho_2\ll\rho_e \,, \\
v\sim \mathrm{constant~}  \ &\mbox{when} \quad \rho_2\gg\rho_e \, .
\end{array}
\right.
\end{align}
Assuming the environmental density has a power-law form of $\rho_e\sim10^{-y}\ [\mathrm{g/cm^3}]$, we can estimate the critical radius $r_c$ for the tangential velocity $v$, given by the condition $3\rho_2=\rho_e$:
\begin{align}
\label{rcn}
r_c\simeq 2\times10^{y/2-5} \, v \ [\mathrm{pc}] \, .
\end{align}
The tangential velocity is usually of ${\cal O}(10^{-3})\sim {\cal O}{(10^{-4})}$ in natural unit, or ${\cal O}(10)\sim {\cal O}{(10^2)}[\mathrm{km/s}]$, according to observations on galaxies (see \cite{Sofue:2000jx,Lelli:2016zqa} for more details). For instance, for $y=23$, namely $\rho_e\sim 10^{-114} M_{p}^4$ while the tangential velocity of test particle $v\simeq3\times10^{-4}$ in the galaxy, we have
\begin{align}
\label{rcondition}
r_c\sim2 \ [\mathrm{kpc}] \, .
\end{align}
Another example is to input $y=17$, $v\simeq3\times10^{-4}$ for the solar system, one can obtain $r_c=1.9 [\mathrm{pc}]$ .
It is acceptable that $r_c$ is much larger than the size of the object as a gravitational source in order to maintain the Newtonian gravity.

Moreover, in the non-relativistic limit, Eq.~\eqref{dmdisk2} implies
\begin{align}
\label{pv}
p-1\simeq v^2/2\,,
\end{align}
and for a typical value of $v^2\sim10^{-6}$, one has
\begin{align}
\label{pv2}
p-1\sim10^{-6}\,.
\end{align}

\begin{figure}[htbp]
\centering
\includegraphics[width=0.6\textwidth]{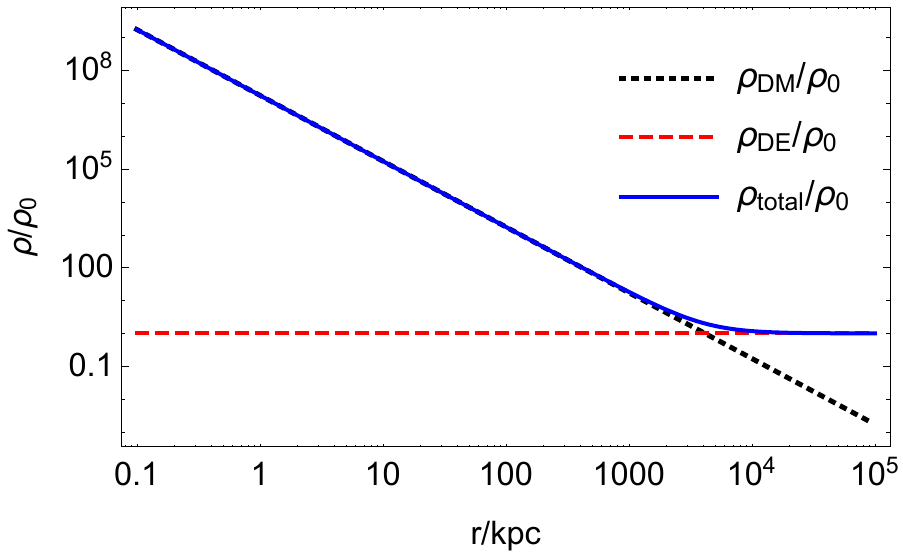}
\caption{
The density profiles of DM, DE, and the total energy with the parameters chosen as $n=-1$, $x\rightarrow\pm\infty$ and $p-1=10^{-6}$.
The energy density is normalized by $\rho_0 \equiv 1/(2p-1)(-z)=\Lambda/\zeta(2p-1)$.
The black dotted line stands for DM density, the red dashed for DE density, and the blue solid for the total density.
}
\label{f3}
\end{figure}

{\it Post-Newtonian parameter.} The post Newtonian parameter $\gamma$ describes how the current theory of gravity deviates from GR. For a massive object whose gravity causes deflection of the light-ray passing it, the deflection angle $\varphi$ is expressed as \cite{Will:2014kxa}:
\begin{align}
\varphi\propto\frac{(1+\gamma)M}{2d}~,
\end{align}
where $\gamma$ is the post-Newtonian parameter, $M$ is the mass of the object, and $d$ is the distance between the center of the gravity potential and the light ray (see Ref.~\cite{Jusufi:2017vta} for detailed analysis based on the Gauss-Bonnet theorem). In our model, we find that $\gamma$ relates with $p$ as:
\begin{align}
\label{dol}
\frac{1+\gamma}{2}=\frac{1}{\sqrt{2p-1}}\,,
\end{align}
The detection of the deflection angle puts severe constraints on $\gamma$ as \cite{Will:2014kxa}:
\begin{align}\label{constriantgamma}
|\gamma-1|<2.3\times10^{-5}.
\end{align}
which implies
\begin{align}
\label{pppn}
|p-1|\simeq\frac{1}{2}|\gamma-1|<1.15\times10^{-5}\,.
\end{align}
The constraints Eq.\eqref{pv2} and Eq.\eqref{pppn} on $p$ given by the two experiments are therefore consistent with each other. 
Fig.~\ref{f3} illustrates the schematic idea of the two-fluid model and displays the density profiles of the DM and DE.

\section{Dark Matter Behavior in Cosmological Evolution}
\label{DMBcos}
In the last sections, we use the model \eqref{model} to generate a spherically symmetric black hole solution, and use the two-fluid description to give rise to the behaviors of DM and DE (in form of cosmological constant) in local frame.
However,  \cite{Casalino:2018mna} shows that our model with $p=1$ breaks the consistency between the DM behavior in cosmological evolution and the constraints on the speed of the gravitational wave.
This inconsistency also happens as $p\simeq1+v^2/2$ in our model. Therefore, we need to improve our model to get consistent predictions with other physics. In the following, we consider a simple extension of our toy model in which we can keep the black hole solution obtained in the previous section.

\subsection{Improved Model for Cosmology}

We choose the new choice of functions in the following form:
\begin{align}
\label{model2}
\begin{split}
G_{2} &= -2\Lambda + 2\eta X^p + 2\alpha(X-X_1)^2(X-X_0)^q
\, , \\
G_{4} &= 1+\beta X^p \, ,
\end{split}
\end{align}
where $G_{2}$ is corrected by the third term $\tilde{G}_{2}(X) \equiv 2\alpha(X-X_1)^2(X-X_0)^q$ where $X_{1}$ is an additional parameter, and $\zeta=1$ is assumed in the original model Eq.~\eqref{model}. The corrected term is inspired by the interesting model dubbed as purely kinetic k-essence \cite{Scherrer:2004au, Bertacca:2010ct} where the Lagrangian is written as $L=-\Lambda+\alpha (X-X_0)^q$ with $q\geq 2$, which is claimed to give an almost pressure-less DM fluid.

Imposing $\beta\Lambda+\eta=0$, we can still obtain the black hole solution in vacuum Eq.~\eqref{FG} when we choose $X_1=\sqrt[p]{(p-1)\Lambda/(2p-1)\eta}$ thanks to $(X-X_1)^2$.
Indeed, the additional term to original $G_2$ function yields to the condition $\tilde{G}_2(X=X_1)=\tilde{G}_{2X}(X=X_1)=0$. $(X-X_0)^q$ makes the scalar field behave like DM in cosmological evolution as in the purely kinetic k-essence. Now we will investigate the cosmological evolution in the improved model Eq.~\eqref{model2}.

In flat FRW metric, $ds^{2}=-dt^{2}+a^{2}(t) \delta_{ij}dx^{i}dx^{j}$ where $a(t)$ is the cosmic scale factor, the Hubble parameter is defined as $H=\dot{a}/a$. The Friedmann equation takes the following form,
\begin{align}
\label{TE}
E&\equiv3H^{2} \nonumber \\
&=\tilde{\rho}_{\phi}+\tilde{\rho}_{r}+\tilde{\rho}_{b}
\nonumber \\
&=\frac{1}{M_{*}^{2}}[-G_{2}+2XG_{2X}+12H^{2}(XG_{4X}+2X^{2}G_{4XX})]+\tilde{\rho}_{r}+\tilde{\rho}_{b} \, ,
\end{align}
where the effective Planck mass squared $M_{*}^{2}$ is
\begin{align}
\label{msq}
M_{*}^{2}=2(G_{4}-2XG_{4X}) \, ,
\end{align}
and $\tilde{\rho}_{r}$, $\tilde{\rho}_{b}$ , $\tilde{\rho}_{\phi}=\tilde{\rho}_{DM}+\tilde{\rho}_{DE}$ respectively denote the induced energy densities of radiation, baryon and scalar field containing DM and the DE, with
\begin{align}
\label{EOMcos5}
\tilde{\rho}_{DE}
&=\Lambda
=3H_{0}^{2}\Omega_{\Lambda0} \, , \\
\label{EOMcos51}
M_*^2\tilde{\rho}_{b}
&=3H_{0}^{2}\Omega_{b0}a^{-3} \, , \\
\label{EOMcos6}
M_*^2\tilde{\rho}_{r}
&=3H_{0}^{2}\Omega_{r0}a^{-4} \, .
\end{align}
Here, we set the scale factor $a_0=1$ at present and $\Omega_{i0}$ is the current value of the density fraction.
Another equation of motion about the total pressure $\tilde{P}$ including the baryon, radiation, and the scalar field is given by
\begin{align}
\label{TP}
\tilde{P}
&\equiv -(3H^{2}+2\dot{H}) \nonumber \\
&=\tilde{p}_{\phi}+\tilde{p}_{r}+\tilde{p}_{b}\nonumber \\
&=\frac{1}{2(G_{4}-2XG_{4X})}(G_{2}-8HX\dot{X}G_{4XX}-4H\dot{X}G_{4X})+\tilde{p}_{r}+\tilde{p}_{b},
\end{align}
where
$\tilde{p}_{b}=0$, $\tilde{p}_{r}=\tilde{\rho}_{r}/3$ and $\tilde{p}_\phi=\tilde{p}_{DM}+\tilde{p}_{DE}$, with $\tilde{p}_{DE} =-\Lambda$.
The equation of motion for scalar field in the FRW Universe is then written as
\begin{align}
\label{scalarcos}
\dot{X}=\frac{3}{2}XH^{-1}\alpha_{K}^{-1}(2\tilde{P}\alpha_{B}-8G_{2X}X) \, ,
\end{align}
where the braiding parameter $\alpha_B$ is defined as
\begin{align}
\label{aB}
\alpha_{B}=\frac{8(XG_{4X}+2X^{2}G_{4XX})}{M_{*}^{2}}  \, ,
\end{align}
and the the kineticity parameter $\alpha_K$ is defined as
\begin{align}
\label{ak}
\alpha_{K}=\frac{1}{M_{*}^{2}H^2}
\left[12H^2(4X^{3}G_{4XXX}+8X^{2}G_{4XX}+XG_{4X})+4X^{2}G_{2XX}+2XG_{2X} \right]  \, .
\end{align}
From Eqs.\eqref{TE} and \eqref{TP}, we define the effective pressure and energy density of scalar DM field
\begin{align}
\label{rhocos}
M_{*}^{2}\tilde{\rho}_{DM}
&=-G_{2}-M_{*}^{2}\Lambda+2XG_{2X}+\frac{3}{2} M_{*}^{2} \alpha_{B} H^{2} \,, \\
\label{pcos}
M_{*}^{2}\tilde{p}_{DM}
&=G_{2}+M_{*}^{2}\Lambda-M_{*}^{2}\alpha_{B}H\frac{\ddot{\phi}}{\dot{\phi}} \, .
\end{align}

In the model \eqref{model2}, when the correction term $\tilde{G}_{2}(X)$ is dominant in the cosmological evolution compared to the $\eta X^p$, we can evaluate the equation of state of DM as
\begin{align}
\label{eos}
w_{DM}\equiv\frac{\tilde{p}_{DM}}{\tilde{\rho}_{DM}}=\frac{(X-X_0)(X-X_1)}{(3+2q)X^2-X_0X_1+X(-3X_0+X_1-2qX_1)}.
\end{align}
In order to solve Eq.\eqref{scalarcos}, we assume $\tilde{P}\alpha_{B}\ll G_{2X}X$ and $ G_{4}\simeq \text{constant}$, which is very reasonable for small $\beta$. Thus we can obtain the following solution \cite{Scherrer:2004au, Bertacca:2010ct}
\begin{align}
\label{Xsolotion}
XG_{2X}^2=ka^{-6},
\end{align}
where $k$ is a positive constant.
Furthermore, if we define $\varepsilon\equiv(X-X_0)/X_0\ll1$, from above equation we have
\begin{align}
\label{e}
\varepsilon=\left(\frac{a}{a_q}\right)^{-3/(q-1)}\,,
\end{align}
where $a_q^{-3}=\alpha(X_1-X_0)^2 q X_0^q(X_0k)^{-1/2}$. And then, the equation of state of the scalar field approximates
\begin{align}
\label{eos1}
w_{DM}\simeq\frac{\varepsilon}{2q}\, ,
\end{align}
and the energy density is evaluated as
\begin{align}
\label{ed1}
\rho_{DM}\equiv M_{*}^{2}\tilde{\rho}_{DM}\simeq 2q\alpha(X_1-X_0)^2 X_0^q \left(\frac{a}{a_q}\right)^{-3} \,.
\end{align}
Because the scalar field should behave like the DM before the epoch of the matter-radiation equality we need the condition $\varepsilon\ll1$ at that epoch. Using Eq.~\eqref{e}, the necessary condition is $a_{eq}\gg a_q$ where $a_{eq}$ is the scale factor at the epoch of the matter-radiation equality, given by $a_{eq}=3\times10^{-4}$. Notice the energy densities of DE and DM in the current universe has a relation $\rho_{DM0}/\Lambda=\Omega_{DM0}/\Omega_{\Lambda0}$. Combining with Eq.~\eqref{ed1}, one can obtain
\begin{align}
\label{aaq}
a_q^3=\frac{\Lambda\Omega_{DM0}}{2q\Omega_{\Lambda0}\alpha(X_1-X_0)^2 X_0^q}\ll a_{eq}^3\simeq 3\times 10^{-11}\,.
\end{align}
To satisfy the above constraint, $\alpha$ can be so large that it makes $\tilde{G}_{2}(X)$ dominated in cosmological evolution. Furthermore, the equation of state of DM has been constrained by observations \cite{Kopp:2018zxp}, i.e. $|w_{DM}|\ll1$, which is surely guaranteed by Eq.~\eqref{eos1}.

Moreover, when $\varepsilon\gg1$, from Eq. \eqref{eos} the equation of state becomes
\begin{align}
\label{eos2}
w_{DM}\simeq\frac{1}{2q+3}\,.
\end{align}
since this case correspond to the very early universe when DM has not become important yet, we will not discuss it anymore.

\subsection{Instability and Speed of Gravitational Wave}

We have discussed the cosmological evolution of the scalar-field DM in the previous subsection, to find that we can make the equation of state parameter $w_{\phi}$ smaller than the unity, which realizes the almost pressure-less DM in the cosmic history.
In the following, we consider other aspects of the scalar field: the sound speed of the scalar field and the speed of the gravitational waves.
The sound speed $c_s$ is related to the instability of scalar perturbation. The sound speed has to be small enough to generate the cosmic large-scale structure formation and CMB temperature anisotropies \cite{Bertacca:2010ct}.
For instance, it should be extremely small $c_s^2< 10^{-10.7}$ at present constrained by observations \cite{Kunz:2016yqy}. Meanwhile, the speed of gravitational wave $c_T$ is also tightly constrained by the observation of GW170817 and GRB 170817A \cite{GBM:2017lvd}.

The action at quadratic order of a scalar field $\zeta$ and tensor modes $h_{ij}$ are given by
\begin{align}
S_{2} = \int{dtd^3xa^3}
\Biggl [Q_s \left(\dot{\zeta}^2-\frac{c^2_s}{a^2}(\partial_i \zeta)^2\right)
+ Q_T \left(\dot{h}_{ij}^2-\frac{c^2_T}{a^2}(\partial_k h_{ij})^2\right)  \Biggr]\,,
\label{S2}
\end{align}
where (see appendix \ref{appendix})
\begin{align}
\label{Qs}
Q_s
&= \frac{2M^2_*D}{(2-\alpha_B)^2} \, , \\
\label{cs2}
c^2_s
&=-\frac{\left(2-\alpha_{\mathrm{B}}\right)\left[\dot{H}-\frac{1}{2} H^{2} \alpha_{\mathrm{B}}\left(1+\alpha_{\mathrm{T}}\right)-H^{2}\left(\alpha_{\mathrm{M}}-\alpha_{\mathrm{T}}\right)\right]-H \dot{\alpha}_{\mathrm{B}}+\tilde{\rho}_{\mathrm{m}}+\tilde{p}_{\mathrm{m}}}{H^{2} D} \, , \\
D & \equiv \alpha_K + \frac{3}{2}\alpha^2_B \,,
\end{align}
and
\begin{align}
\label{aT_appG24}
& M^{2}_{*} \alpha_{T} \equiv 4XG_{4X}
\,,\\
\label{aM_appG24}
&HM^2_*\alpha_M \equiv \frac{d}{dt}M^2_{*}
\,, \\
\label{QTG24}
&Q_T = \frac{M^2_*}{8}
\,, \\
\label{cT2G24}
&c^2_T = 1 + \alpha_{T} \,.
\end{align}
Here $\alpha_M$ is the rate of running of the Planck mass and $\alpha_T$ is the tensor speed excess. To avoid the theoretical instabilities, we should impose the conditions $Q_s > 0$, $c^2_s > 0$, $Q_T > 0$ and $c^2_T >0$. If the tensor speed excess is small $\alpha_T\ll1$ and $\alpha_M,\alpha_B,\dot{\alpha}_B\ll1$, namely $G_4$ is almost a constant, the speed of the scalar perturbation mode is given by
\begin{align}
\label{cs2cg}
c^2_s \simeq \frac{1}{H^2\alpha_K} (\tilde{\rho}_\phi+\tilde{p}_\phi ) \,,
\end{align}
utilizing  Eq.\eqref{eos} and $\varepsilon\ll1$, the speed of scalar perturbation is given by
\begin{align}\label{cse}
c_s^2\simeq \frac{\varepsilon}{2(q-1)} \, ,
\end{align}
thus we have a small sound speed of scalar perturbation which behaves like the DM.

For the tensor perturbation, from Eq.~\eqref{aT_appG24}, the deviation of the speed of gravitational wave from that of light in the low-redshift era can be evaluated as
\begin{align}
\label{gwlight}
\alpha_T\simeq\frac{2p\beta X_0^p}{1-(2p-1)\beta X_0^p}
\end{align}
According to the observation constraint from GW170817 and GRB 170817A, the difference between speeds of the gravitational wave and light is should be suppressed \cite{GBM:2017lvd},
\begin{align}\label{atobs}
-3\times 10^{-15}<c_T-1<7\times 10^{-16}\, .
\end{align}
Thus, by combining Eqs.~\eqref{cT2G24}, \eqref{gwlight}, and \eqref{atobs}, $p\simeq1$ and $\beta<0$ gives the constraint on $X_{0}$:
\begin{equation}
\label{bXp}
-3\times 10^{-15}<\beta X_0<0 \,.
\end{equation}


\section{Conclusion and Discussion}
\label{sec:conclusion}
We have investigated the shift-symmetric subclass of the Horndeski theory, where $G_{2}$ and $G_{4}$ are the power functions of the kinetic term $X^{p}$ as in Eq.~\eqref{model},
to find the spherically symmetric solutions with the vanishing Noether current of the scalar field.
We have also found that the scalar-field fluid becomes anisotropic for $p\neq 1$.
Based on the two-fluid model,
we have decomposed the scalar field fluid into two parts corresponding to the DM and DE.
We have found the DM part of energy density scales as the inverse square of the radial coordinate
in the spherically symmetric solution which we have derived,
and we can explain the observed galaxy rotation curve in our scenario.

Moreover, by calculating the gravitational mass $m(r)$ including both baryonic matter and scalar field,
we have confirmed that the constant velocity appears in the galaxy rotation curve at $\mathcal{O}(\mathrm{kpc})$ scale
and that the parameter $p$ is related to the velocity $v$ by $p\simeq1+v^2/2$ in the non-relativistic limit.
Using the observational constraint on the velocity, $v\simeq{\cal O}(10^{-3})\sim {\cal O}{(10^{-4})}$,
we can put constraint on parameter $p$.
Furthermore, $p$ can also be related to the post-Netwonian parameter $\gamma$ in terms of $p\simeq (3-\gamma)/2$,
where $|\gamma-1|<2.3\times10^{-5}$ is given by the deflect angle of light passing the massive object.
One can find that the two constraints on $p$ are consistent with each other.

Since the above model \eqref{model} suffers from the inconsistency between the cosmic evolution and the gravitational wave speed,
we have studied the correction term $\alpha(X-X_1)^2(X-X_0)^q$ added to this model while keeping the black hole solution.
Based on this corrected model, we have evaluated the effective energy density and pressure of the scalar field,
showing that the corrected model gives a DM behavior in cosmic history when the additional term is dominant and the constraint Eq.~\eqref{aaq} is satisfied.
Regarding the speed of the gravitational wave, although it is not exactly equal to the speed of light,
we can obtain the consistent result with the observation if $-3\times 10^{-15}<\beta X_0<0$.
The constraint Eq.~\eqref{aaq} and Eq.~\eqref{bXp} can be consistent with each other thanks to the new parameter $\alpha$.

Before closing, let us make several comments for future works.
First of all, we have assumed a specific combination of the parameters $\beta \Lambda + \zeta \eta = 0$ for the simplicity, 
and our analysis has been demonstrated in a specific black hole solution.
It would be interesting to investigate the other parameter regions to keep $X$ or $X^{p}$ analytically well-defined.
Moreover, regarding the two-fluid model,
we have observed the product of 4-velocities of two fluids $K = v_{\mu} u^{\mu} \rightarrow - \infty$ in our calculation.
The above divergence shows up because we have assumed cosmological constant $w = -1$ for the DE fluid.
Thus, if we admit a slight change of the equation-of-state parameter,
that is, a small deviation from the cosmological constant for the DE,
$K$ does not diverge, and the relative speed is finite.

In relation to the equation of state of the DE,
it is not necessary to choose the cosmological constant for the DE part in the two-fluid model although the present work focuses on the simplest case
which induces the dark side of the Universe in a way similar to the $\Lambda$CDM model.
We can generalize the current analysis by changing the parameter $n$,
which characterizes the equation-of-state parameter of the DE.
If the DE behaves as the phantom in our model,
the profile of the DM density is shifted by the constant term as in Eq.~\eqref{density_profiles}
although such a constant is of the order of the DE scale.

Furthermore, the density profiles of two dark components also depend on Eq.~\eqref{Tmunu0}. 
In our study, we have identified the roles of the dark sectors from each component of the effective energy-momentum tensor as Eqs.~\eqref{rhop}-\eqref{rhop3} based on our spacetime solution Eqs.~\eqref{FG}-\eqref{FG3}; 
that is, the constant part corresponds to the cosmological constant, and the inverse-square part does to the DM. 
Thus, a new spacetime solution could allow us to consider the nontrivial density profiles, 
and we could realize the quintessential DE in such a new solution although we only found the cosmological constant
and phantom DE in the present analysis.
Such a new solution might also realize the density profile of the DM which deviates from the simple inverse-square law.

It is also of great importance to reconsider our assumptions about the symmetry and model-building.
For instance, we can generalize our work to include time-dependence for the scalar field $\phi$ as in \cite{Kobayashi:2014eva} and to refine our study on the galaxy rotation curve in the stationary spacetime solution.
Regarding the constraints on the parameter from the speed of gravitational waves,
it would be milder if we invoke the disformal transformation $\tilde{g}_{\mu \nu}=g_{\mu \nu}+D(X, \phi) \phi_{\mu} \phi_{\nu}$ with an appropriate expression of function $D(X, \phi)$ \cite{Babichev:2017lmw,Ezquiaga:2017ekz}.
By starting from the study on the scalar-field DM demonstrated in this paper,
it is necessary to apply it to the other sub-classes of the Horndeski theory or beyond-Horndeski theories.


\begin{acknowledgments}

T.K. is supported by International Postdoctoral Exchange Fellowship Program at Central China Normal University and Project funded by China Postdoctoral Science Foundation 2018M632895.
T.Q. is supported by the National Science Foundation of China (NSFC) under Grant No.~11875141 and No.~11653002.

\end{acknowledgments}


\appendix

\section{Cosmological evolution and perturbation analysis in Horndeski theoty}
\label{appendix}
In Sec. \ref{DMBcos}, we follow the $\alpha$ parametrization introduced by Bellini and Sawacki \cite{Bellini:2014fua,Arai:2017hxj}, which is given as follows:
\begin{align}
&M^2_* \equiv 2 (G_4 -2XG_{4X} + XG_{5\phi} - \dot{\phi}HXG_{5X})\,,\\
&HM^2_*\alpha_M \equiv \frac{d}{dt}M^2_*\,, \label{aM_app}\\
&H^2M^2_*\alpha_K \equiv 2X(G_{2X}+2XG_{2XX}-2G_{3\phi}-2XG_{3\phi X}) \cr
&\qquad \qquad +12\dot{\phi}XH(G_{3X}+XG_{3XX}-3G_{4\phi X}-2XG_{4\phi XX}) \cr
&\qquad \qquad +12XH^2(G_{4X}+8XG_{4XX}+4X^2G_{4XXX}) \cr
&\qquad \qquad -12XH^2(G_{5\phi}+5XG_{5\phi X}+2X^2G_{5 XXX}) \cr
&\qquad \qquad + 4\dot{\phi}XH^3(3G_{5X}+7XG_{5XX}+2X^2G_{5XXX})\,, \label{aK_app} \\
&HM^2_*\alpha_B \equiv 2\dot{\phi}(XG_{3X}-G_{4\phi}-2XG_{4\phi X})\cr
&\qquad \qquad + 8XH(G_{4X}+2XG_{4XX}-G_{5\phi} -XG_{5\phi X}) \cr
&\qquad \qquad + 2\dot{\phi}XH^2(3G_{5X}+2XG_{5XX})\,, \label{aB_app}\\
&M^2_*\alpha_T \equiv 2X(2G_{4X} - 2G_{5\phi} - (\ddot{\phi} - \dot{\phi} H)G_{5X})\,\label{aT_app}.
\end{align}
The the Friedman equations in the Horndeski theory are given by
\begin{align}
&3H^2 = \tilde{\rho}_m + \tilde{\rho}_\phi\,, \label{F1}\\
&2\dot{H} + 3H^2 = -\tilde{p}_m - \tilde{p}_\phi\,, \label{F2}
\end{align}
where $\tilde{\rho}_m \equiv \rho_m/M^2_*$ and $\tilde{p}_{m} \equiv p_m/M^2_*$. Then $\tilde{\rho}_\phi$ and $\tilde{p}_\phi$ are given by
\begin{align}
M^2_*\tilde{\rho}_\phi
&= -G_2+2X(G_{2X}-G_{3\phi}) \cr
&+6\dot{\phi} H(XG_{3X}-G_{4\phi}-2XG_{4\phi X})\cr
&+12H^2X (G_{4X}+2XG_{4XX}-G_{5\phi}-XG_{5\phi X})\cr
&+4\dot{\phi}H^3X(G_{5X}+XG_{5XX})\,, \label{E}\\
M^2_*\tilde{p}_\phi
&= G_2-2X(G_{3\phi}-2G_{4\phi\phi})\cr
&+4\dot{\phi}H(G_{4\phi}-2XG_{4\phi X} + XG_{5\phi\phi})\cr
&-M^2_*\alpha_B H\frac{\ddot{\phi}}{\dot{\phi}}-4H^2X^2G_{5\phi X} + 2\dot{\phi}H^3XG_{5X}\,.\label{P}
\end{align}

The action at quadratic order of a scalar field $\zeta$ and tensor modes $h_{ij}$ are given by
\begin{align}
&S_2 = \int{dtd^3xa^3}
\Biggl [Q_s \left(\dot{\zeta}^2-\frac{c^2_s}{a^2}(\partial_i \zeta)^2\right)
+ Q_T \left(\dot{h}_{ij}^2-\frac{c^2_T}{a^2}(\partial_k h_{ij})^2\right)  \Biggr]\,,
\end{align}
where
\begin{align}
&Q_s = \frac{2M^2_*D}{(2-\alpha_B)^2}, \\
&c^2_s = -\frac{\left(2-\alpha_{\mathrm{B}}\right)\left[\dot{H}-\frac{1}{2} H^{2} \alpha_{\mathrm{B}}\left(1+\alpha_{\mathrm{T}}\right)-H^{2}\left(\alpha_{\mathrm{M}}-\alpha_{\mathrm{T}}\right)\right]-H \dot{\alpha}_{\mathrm{B}}+\tilde{\rho}_{\mathrm{m}}+\tilde{p}_{\mathrm{m}}}{H^{2} D} \,,\\
&D \equiv \alpha_K + \frac{3}{2}\alpha^2_B\,,\nonumber
\end{align}
while
\begin{align}
&Q_T = \frac{M^2_*}{8}\,,\label{QT}\\
&c^2_T = 1+\alpha_T\,,\label{cT2}
\end{align}
To avoid the ghost and gradient instability, we should impose the condition that $Q_s > 0$, $c^2_s > 0$, $Q_T > 0$ and $c^2_T >0$.

\bibliographystyle{apsrev4-1}
\bibliography{HDM2019}

\end{document}